# Effect of grooves and checkerboard patterns on the electron emission yield

J. Pierron[1,a], C. Inguimbert[1,a], M. Belhaj[1,a], J. Puech[2,b], and M. Raine[3,c]

[1]*ONERA, The French Aerospace Lab, 2 Avenue Edouard Belin, 31055 Toulouse, France*

[2]*CNES, 18 Avenue Edouard Belin, 31401 Toulouse, France*

[3]*CEA, DAM, DIM, F-91297 Arpajon, France*

**ABSTRACT**

The effect of rough structures on the electron emission under electron impact between 10 eV and 2 keV is investigated with a new version of the low energy electromagnetic model of GEANT4 (MicroElec). The inelastic scattering, is modeled thanks to the dielectric function theory and the Mott's model of partial waves to describe the elastic scattering. Secondary electron emission is modeled for grooved and checkerboard patterns of different dimensions for aluminum and silver. The analyses is performed according to two shape parameters h/L and d/L, h being the height, L the width of the structures and d the spacing between two neighboring structures. The secondary electron emission is demonstrated to decrease when h/L and d/L ratios increase. When the height reaches 10 times the lateral dimensions, the electron emission yield is divided by two compared to that of a flat sample. The optimization of the two aspects ratios lead to a reduction of the electron emission yield of 80 % for grooved patterns, and of 98 % for checkerboard patterns. This purely geometric effect is similar for aluminum and silver materials. A simple analytical model, capable to reproduce the effect on the electron emission yield of checkerboard and grooved patterns, is proposed. This model is found to be in good agreement with the Monte Carlo simulations and some experimental measurements performed in our irradiation facility.

## I. INTRODUCTION

Low energy electrons of a few tens and up tp some hundreds of eV, under the action of the RF electric field, may cause Multipactor breakdowns in satellite communication devices such as microwave and millimeter waveguide equipment[1,2]. The triggering power threshold of these vacuum discharges strongly depends on the Electron Emission Yield (EEY) of the walls of the component, i.e. the number of electrons extracted from the walls under the impact of an incident electron. If the EEY is greater than 1, undesirable electron-clouds can be produced. If the impact of the electron-clouds with the walls of the device are synchronized with the RF electric field , it can initiate a resonant effect that can lead to the triggering of a Multipactor discharge. These discharges may degrade the component per-formance, such as disrupting the transmitted signal or physically damaging the walls. To prevent Multipactor breakdowns, materials with a low EEY are required. However, many materials, including those used for space applications[2], have an EEY greater than 1 for incident electrons below 2 keV. At such low energies, the main issue is that the electrons tend to remain within the first ten nanometers from the surface[3]. As a consequence, the surface properties of the sample such as hydrocarbon contamination[4], oxidation[5] or roughness[5,6] may have a great influence on the EEY. In many different fields, engineered materials with specific surface state are developed to mitigate the SEY phenomenon. Materials with typical surface roughness, known to reduce SEY by trapping the secondary electrons is an option increasingly used. Both experimental and numerical works have been recently been published in this field[7-17]. The present work is in line with those approaches. Our work proposes to study the effect of roughness on the EEY at low energy (10 eV – 10 keV). Depending on the shape of the structures on the surface of the material, the roughness can have two opposite effects. First, the morphology of the surface increases the mean angle of incidence of primary electrons, and therefore may increase the EEY[6]. Second, the surface asperities can act as traps for emitted electrons that can be re-captured by the solid[14]. These shadowing effects are dominant in materials with high roughness[15]. Some authors[15-17] have worked on lowering the EEY by creating surface asperities chemically[15], mechanically[16] or using laser engraving[17]. In this paper, we study the effect of grooves and checkerboard patterns by means of 3D Monte Carlo simulations. The MicroElec[18–19] module of low electromagnetic physics of the code GEANT4[20-21] has been modified to allow the simulation of the transport of the electrons down to few eV for three materials (Si, Ag, Al)[22–23]. Thanks to this model the electron transport is modelled in geometries based on different roughness patterns, and the Electron emission Yield (EEY) is studied as a function of different geometrical parameters. An analytical expression is also proposed to evaluate the impact of the roughness on the EEY. Some comparisons with experimental measurements performed on samples having different roughness patterns are also shown. It depicts a good agreement with the Monte Carlo modelling.

[a] C. Inguimbert and M. Belhaj are with ONERA, 2 av. E. Belin, 31055 Toulouse, France (tel: 33-562252734, email: Christophe.Inguimbert@onera.fr , Mohamed.Belhaj@onera.fr).
[b] J. Puech is with CNES, 18 av. E. Belin, 31401 Toulouse, France (email: Jerome.Puech@cnes.fr )
[c] M. Raine is with CEA, DAM, DIF, F-91297 Arpajon, France (email: Melanie.Raine@cea.fr)

## II.    MONTE CARLO CODE FOR LOW ENERGY ELECTRONS

In this work, 3D simulations of both silver and aluminium surface with different rough patterns have been performed with the electromagnetic model MicroElec of the code GEANT4[18–21]. Recently, the module MicroElec has been improved to perform the electron transport down to a few eV[22,23] and extended from silicon to aluminum and silver for the needs of RF applications. The Monte Carlo open-access particle-matter interaction code GEANT4 is a powerful tool for 3D simulations[18, 19]. It enables modelling a geometry that can be fully parametrized by the user. For example, the surface roughness may be modeled using a substrate capped with four smaller volumes. The four volumes are designed to create regular grooves or checkerboard patterns (see section IV and V). In our simulations, a monokinetic electron beam impacts the sample with a normal incidence. The beam is set so that it impacts several patterns. To count the number of emitted electrons, the sample is located inside a spherical detector.

The Monte Carlo procedure of the module MicroElec is a standard one. The transport is made step by step unlike the "condensed history" approach used at higher energy (>keV). At each step, the type of interaction the distance traveled between collisions, and the characteristics of the interaction (scattering angles, energy transfers) are randomly selected.. Paths and energy losses are calculated until the electron is emitted out of the solid or until its energy falls under an energy threshold. Since our interest here concerns only the electrons that escape, the energy cutoff for electrons in a solid has been chosen equal to its work function. For aluminum and silver, this energy is of 4.25 eV and of 4.3 eV respectively (relatively to the Fermi level).

The physical models of the new version of MicroElec have been compared to the code OSMOSEE[23–26]. At low energy, the trajectory of electrons in matter is driven by two main mechanisms: the elastic (i.e. deflection by nuclei) and the inelastic scattering (i.e. collisions with electrons). To model these interactions, the code MicroElec requires the total and differential cross sections for each of these interactions. In this work, the elastic cross sections have been obtained using the code ELSEPA[27] from Salvat and coworkers which is based on Mott's model (or partial waves). The inelastic cross sections have been calculated using the complex dielectric function theory as described in Refs. [20, 21, 23, 25–27] and the references therein. To extend the code MicroElec at lower energies, the inelastic cross sections have been calculated down to a few eV. In addition, a model for the crossing of the surface potential barrier has been implemented. This model, which considers an exponential potential barrier, has been described elsewhere[22, 24]. Without a model for the surface potential barrier, our simulation results show that the EEY is overestimated below 10 keV by a factor 2 for aluminum and a factor 1.5 for silver[22]. A more complete description of these codes can be found in refs.[23–25].

## III.    PRELIMINARY STUDIES ON FLAT SAMPLES

### Experimental facility

All measurements were performed in the DEESSE4-6 facility at ONERA. This facility is entirely dedicated and designed to the study of electron emission. A dry turbo-molecular pump associated with an oil-free primary pump allows the system to be maintained at a vacuum level down to 2×10-8 mbar. The tank is grounded. The sample holder allows the variation of the electron incidence angle from 0° (normal incidence angle) to 60°. An ELG-2 electron gun from Kimball Instrument was used. The electron beam was pulsed during EEY measurements to limit the radiation induced surface modification effect. The sample was negatively biased to -9V during EEY to avoid low energy secondary electrons to be recollected by the sample. Prior to measurements, the tank was baked to 180°C for 16 hours. In order to measure the EEY the deposited charge per pulse (ΔQb) is first record for each incident electron energy The sample is thereafter negatively biased (-9 V), to prevent the electrons recollection by the sample. The charge per pulse passing through the sample is monitored (ΔQs). The EEY can be obtained using the following relation expression.

$$EEY = \frac{\Delta Qb - \Delta Qs}{\Delta Qb} \qquad (1)$$

**Secondary emission yields on flat samples**

Before investigating the effect of roughness on the EEY, the MC simulations have been validated by comparisons with experimental EEY measured on bulk samples that were Ar-etched in ultrahigh vacuum with the DEESSE facility. A description of this facility, entirely dedicated to and designed for electron emission measurements, can be found in refs. [4]–[6]. Our previous work[22, 23, 25, 26] shows that for aluminum and silver a relatively good agreement is found between MicroElec simulations, the experimental data from the DEESSE facility made by Gineste[5], and the data of Bronstein *et al.* [28]. The maximum EEY $\sigma$ and the energy of the crossover points $Ec_1$ and $Ec_2$ (energy for which the EEY is equal to 1) are summarized in Table 1 for a flat sample of silver.

Table 1. Value for the maximum of the EEY and for the energy of the crossover points for a flat sample of silver at a normal angle of incidence ($\theta = 0°$) and at $\theta = 60°$.

| Ag | | MicroElec [15]–[16] | Bronstein [21] | Gineste [5] |
|---|---|---|---|---|
| | $\sigma$ | 1.809 | 1.68 | 1.67 |
| $\theta = 0°$ | $Ec_1$ (eV) | 125.7 | 150 | 140 |
| | $Ec_2$ (eV) | 3633.6 | 3400 | - |
| | $\sigma$ | 2.54 | 2.35 | 2.09 |
| $\theta = 60°$ | $Ec_1$ (eV) | 80 | - | 115 |
| | $Ec_2$ (eV) | | - | - |

At a normal incidence ($\theta = 0°$), the overall discrepancy between the EEY is of less than 19 % for aluminum and of less than 21 % for silver (8% above 50 eV)[22]. We can notice that MicroElec simulations slightly overestimate the EEY compared to the measurements of Gineste[5] and of Bronstein *et al.* [28]. At an angle of incidence of $\theta = 60°$, the discrepancy between the three values slightly increases. For silver, the discrepancy between MicroElec simulations and the data of Bronstein *et al.* is of less than 16 %. However, the discrepancy between MicroElec simulations and the measurements of Gineste is of 30 % in average. The discrepancy between the two sets of experimental data (of nearly 15 %) may mainly be explained by the fact that Bronstein *et al.*'s data are measurements on vacuum evaporated samples whereas Gineste's data are measurements on bulk samples that were Ar-etched in ultrahigh vacuum. In addition, it seems that, in the case of Gineste's measurements for silver, the etching process may have changed the surface roughness of the sample[5]. Indeed, the Ar etching induced surface morphology change is a well-known phenomenon[29] The main issue is that at low energy, the EEY strongly depends on the surface properties of the sample, and as a consequence, the data from the literature are scarce and very subject to a relatively high degree of uncertainty[30]. In that sense, we consider that MicroElec simulations for flat surfaces are in relatively good agreement with Bronstein *et al*'s data and with Gineste's measurements.

## IV.     STUDY OF GROOVED PATTERNS

The grooved patterns are described by three parameters (see Fig. 1): the depth $h$ of the groove (or height of the structure), the width $L$ of the structure and the distance d between two neighboring structures.

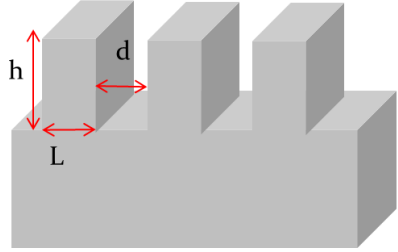

FIG. 1. Illustration of the grooved patterns used in this study.

Some configurations with an optimized set of parameters ($h$, $L$ and $d$) might reduce the EEY. For example, in the case of a pattern with a shallow groove, a very thin groove or a very large groove, the electrons are less likely to be recollected by the neighboring structures: in the first and third configurations, the electrons do not "see" the structures; in the second configuration, the electrons may only hit the top of the structures. Therefore, the EEY should be very similar to that of a flat sample. Contrary to the previous configurations, a pattern with a very deep groove is more likely to decrease the EEY because of the electrons emitted from the grooves that are then recollected by the neighboring structures (shadowing effects). Given these considerations with the geometry, in order to lower the EEY, the effect of the height h of the structures is first investigated. In a second part, the effect of the width $L$ of the structures is studied.

## A. HEIGHT OF THE STRUCTURES

To investigate the effect of the height h of the structures on the EEY, the two parameters L and d are kept constant and of the same value ($L=d$). Fig. 2 shows the simulation results for the EEY of silver with different h values between 0 (flat surface) and 1 mm and with $L=d=50\ \mu m$. For a small height of $h=30\ \mu m$, the EEY is very similar to that of a flat sample (black curve). Then, when the height h increases, the EEY decreases. It can be seen on the figure that the EEY obtained with a height of h=500 μm, 750 μm and 1 mm, present similar values. This behavior is highlighted in Fig. 3 which presents the maximum EEY of the curves in Fig. 2 as a function of the shape factor $h/L$.

For a shape factor greater than 10, it appears that the EEY reaches a plateau of value $\sigma=0.95$. This result can be explained by the fact that when the height is significant compared to the two other dimensions, most of the secondary electrons emitted from the bottom of the structures are recollected by shadowing effect. According to Fig. 3, a shape factor of 10 is sufficient to reduce the EEY of silver by nearly 50 %. This reduction is also obtained for dimensions with the same ratios but a different scale factor (typical dimensions in the range [1μm, 1mm]). In a previous work, a similar result for aluminum has been found [22]. This confirms that the effect of roughness is purely geometrical. It results that, for simple structures such as grooves, a comparison can be made with an analytical model (see section IV. C). Fig. 3 shows that MicroElec simulations are in relatively good agreement with this model (black line). The same study is performed for the backscattered electrons but is not shown here since the results are very similar to the total EEY.

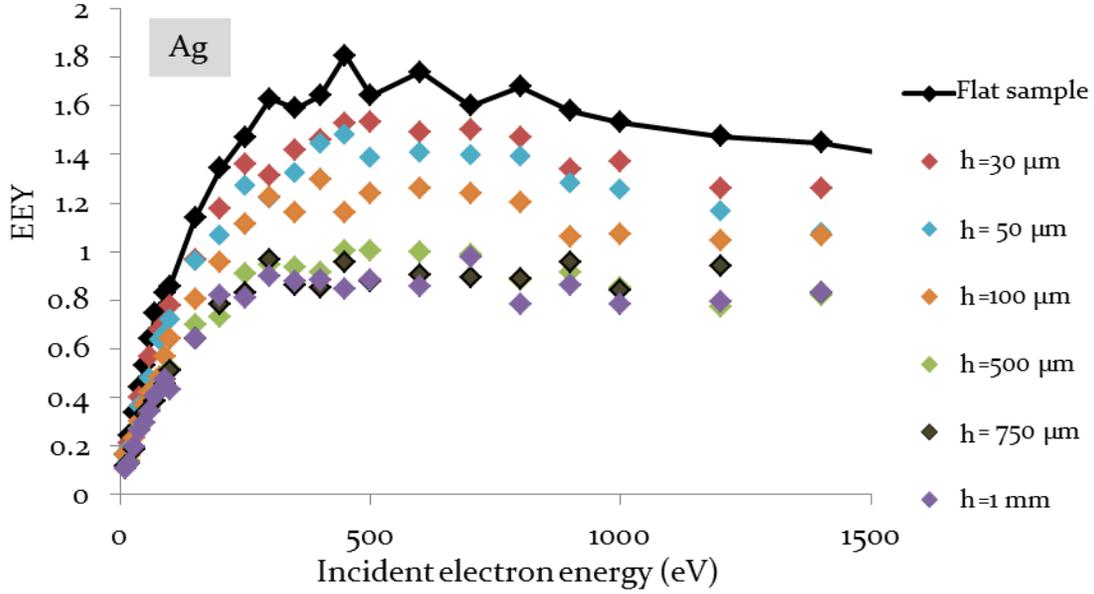

FIG. 2. EEY for a sample of silver with grooved patterns of different height h, width L=50 μm and spacing d=50 μm. The fluctuations correspond to MC uncertainties. For energies below 100 eV, the corresponding error in the EEY is of 10%. For higher energies, the error is of less than 4%. The black curve represents the EEY for a sample with a flat surface.

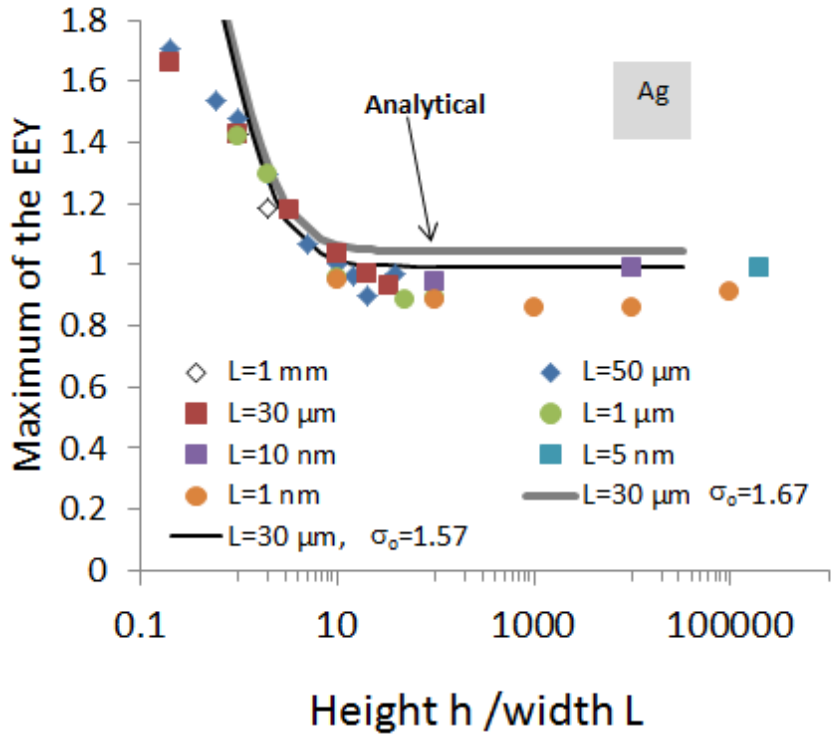

FIG. 3. Maximum of the EEY for a sample of silver with grooved patterns of different height h, width L=50 μm and spacing d=50 μm. The values are calculated with an energy step of 50 or 100 eV for an incident energy between 100 and 600 eV. Uncertainties for the maximum EEY are of less than 5 %. The black and grey lines correspond to the analytical model described in section IV. C for two values of the maximum of EEY ($\sigma_0$ = 1.57 and $\sigma_0$ = 1.67).

To prevent the Multipactor effect, it is also interesting to have a look at the first and second crossover points. As presented in Table 1, MC simulations with the module MicroElec for a flat surface show a value of 125.7 eV for the first crossover point and of nearly 3633.8 eV for the second crossover point. When the EEY decreases, as shown in Fig. 4a, the first crossover point $Ec_1$ is shifted to a higher energy. For example, for a ratio $h/L$=10 with $L$=50$\mu m$, a value of 448.35 eV has been found. On the contrary, as shown in Fig. 4b, the second crossover point $Ec_2$ is shifted to a lower energy. For a ratio $h/L$=10 and $L$=50$\mu m$, a value of 533.1 eV is found. Taking into account the uncertainties of the MC simulations, above a limit value of $h/L$=15, crossover points cannot be observed (EEY<1).

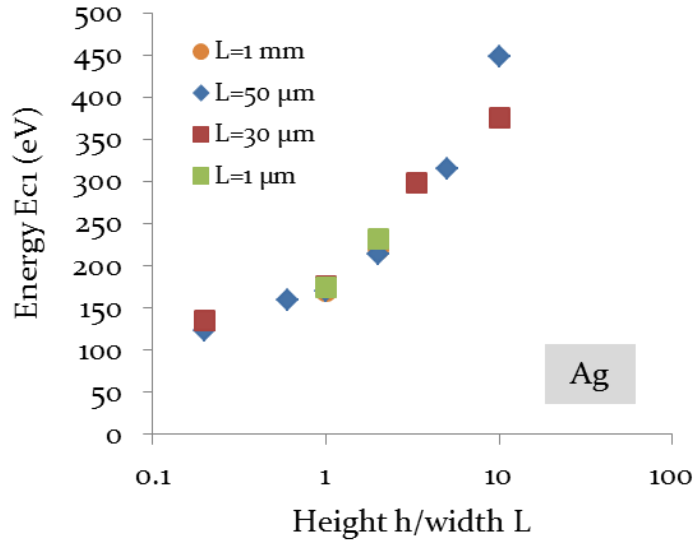

(a)

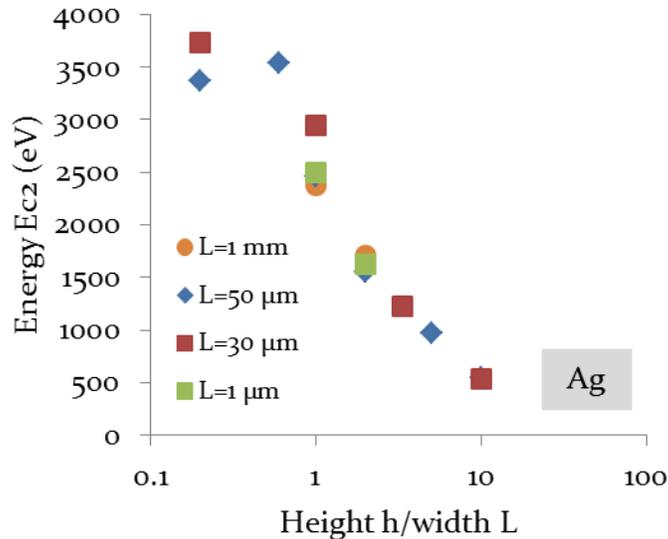

(b)

FIG. 4. Variation of the energy for the (a) first and (b) second crossover points as a function of the ratio h/L with h the height, and L the width of the structures in the case of grooved patterns. The energy of the crossover points is calculated by linear interpolation on the EEY of Fig. 2. The EEY are obtained with a step of 50 or 100 eV between 100 and 600 eV and a step of 200 eV for higher energies.

## B. WIDTH OF THE STRUCTURES

To study the effect of the width *L* of the structures, the two parameters *h* and d are kept constant with *h=10d*. Fig. 5 shows the simulation results for the EEY of silver with different values of widths L between *50 μm* (previous result) and 5 μm, with *d=50 μm* and *h=500 μm*. As can be seen in the figure, when the width L of the structure decreases, the EEY decreases and its shape flattens. The reduction of the yield is highlighted in Fig. 6 which shows the maximum EEY as a function of the ratio of the widths L/d.

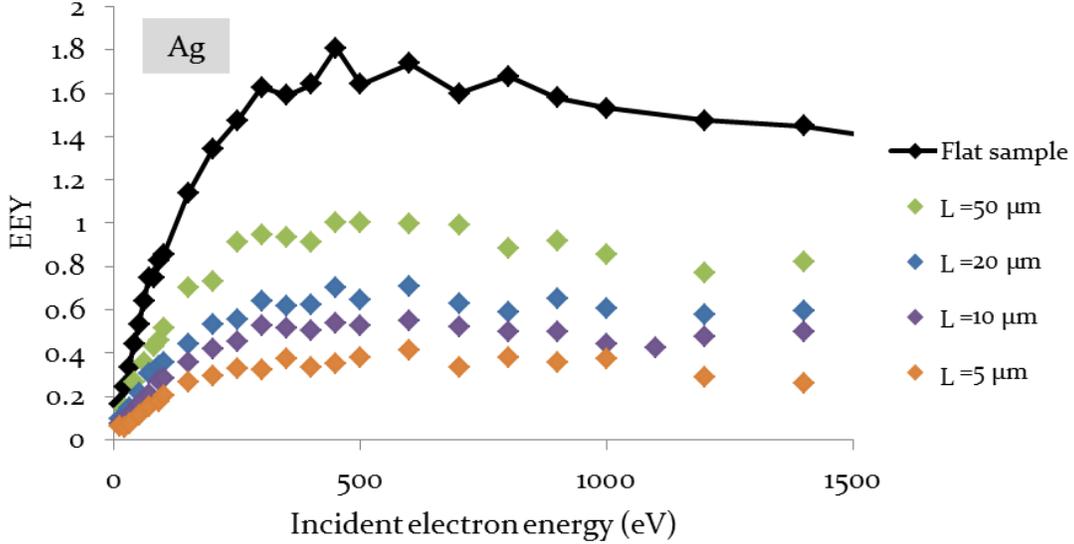

FIG. 5. EEY for a sample of silver with grooved patterns of height h=500 μm, different widths between 1 nm and 1 mm, and spacing d=50 μm. The fluctuations correspond to MC uncertainties. For energies below 100 eV, the corresponding error in the EEY is of 12%. For higher energies, the error is of less than 6%. The black curve represents the EEY for a sample with a flat surface.

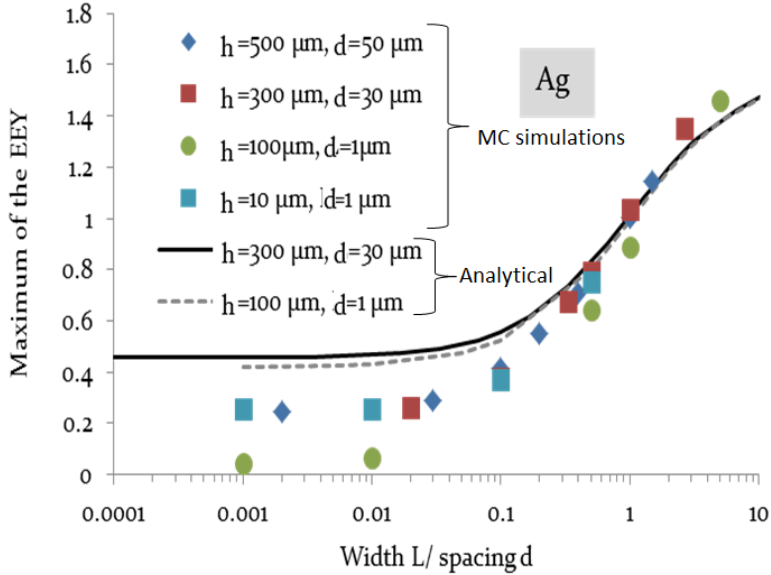

FIG. 6. Maximum of the EEY for a sample of silver with grooved patterns of different height h, different widths L and spacing d. The values are calculated with an energy step of 50 or 100 eV for incident energies between 100 and 600 eV. Uncertainties for the maximum EEY are of nearly 15 % for a low ratio L/d and of nearly 10 % for a ratio L/d>0.01. The black and grey lines correspond to the analytical model described in section IV.C.

For *L/d* smaller than 0.01, the EEY of silver is reduced by nearly 80 % compared to that of a flat sample — the value for the maximum is $\sigma = 0.35$.. This behavior is also depicted in Fig. 6 for *d=30 µm* (*h=300 µm*) and for *d=1µm* (*h=100 µm*). As shown in the figure, the structures with $h/d = 100$ have a lower EEY than structures with $h/d = 10$. As the effect of roughness is purely geometrical, one can see that similar results are obtained for structures with similar ratios but a different scale factor.

## C. ANALYTICAL MODEL

Recently[7,10], analytical models have been proposed to characterize the SEY as a function of the aspect ratio of porous surfaces i. e. a flat surfaces perforated with regularly spaced holes of a given depth. This section describes a similar approach for a grooved pattern, whose results have been presented in Fig. 3 and Fig. 6.According to this approach, the total EEY is considered to be the sum of the electrons emitted from the top of the structures (which we note $\sigma_{top}$), and the electrons emitted from the bottom of the structures that are not recollected by shadowing effects ($\sigma_{bottom\ not\ recollected}$). Neglecting the electrons that are emitted by the edges of the sides of the structures (normal incidence), we obtain the following relation:

$$\sigma = \sigma_{top} + \sigma_{bottom\ not\ recollected} \tag{2}$$

The electrons emitted by the surface of the material have straightline trajectories in the vacuum and can be recollected only by interactions with the walls of the roughness structure. Most of the emitted secondary electrons have a sufficiently low energy to assume that they are trapped following their first interaction, without any reemission of a second generation secondary electron. This phenomenon is thus supposed to be independent of the incident electron energy. This hypothesis is in agreement with the EEYs presented on Fig. 2 and Fig. 5. The number of electrons emitted from the top of the structures, and the number of electrons emitted from the bottom of the structures (before recollection) are directly proportional to the EEY of a flat surface (which we note $\sigma_0$) and to the area irradiated by the incident electrons. Hence, the corresponding yields are given by the relations (2) and (3).

$$\sigma_{top} = \sigma_0 \frac{L}{L+d} \tag{3}$$

$$\sigma_{bottom} = \sigma_0 \frac{d}{L+d} \tag{4}$$

To obtain a general expression for the EEY, in the case of grooved patterns, the number of electrons emitted from the bottom of the structures and then recollected shall be estimated. Fig. 7 shows a view from the top and from the front of the patterns. The electrons that are not recollected are those with an angle θ greater than a limit angle $\theta_{lim}$ and an angle $\varphi$ that can take all the values except $[-\frac{\pi}{2} - \varphi_{lim}; -\frac{\pi}{2} + \varphi_{lim}] \cup [\frac{\pi}{2} - \varphi_{lim}; \frac{\pi}{2} + \varphi_{lim}]$. As a first approach, $\varphi_{lim} = 0$ has been considered. This approximation seems coherent when the grooves are very long; however it overestimates the EEY for a small value of the height h. By considering that in average, the electrons are located in the middle of the grooves at $x=d/2$, an approximated value for the limit angle $\theta_{lim}$ is given by the relation (4).

$$tan\theta_{lim} = \frac{2h}{d}\frac{1}{\sqrt{1 + tan^2\varphi}} \tag{5}$$

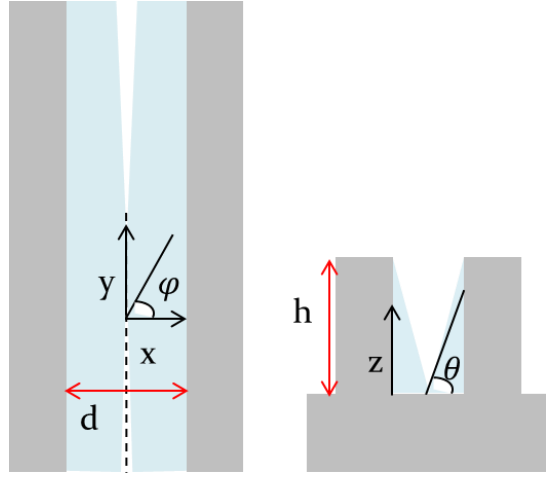

FIG. 7. Illustration of the grooved patterns used in the study. View from the top (left) and from the front (right). The electrons in the blue area are recollected by the structures.

Some Monte Carlo simulations have shown that the angular distribution of emitted electrons follows a Lambertian distribution. Thus assuming a Lambertian distribution for secondary electrons, the yield $\sigma_{bottom\ not\ recollected}$ is given by the equation (5) that can be numerically integrated using a limit angle $\beta_{lim} = \pi/2 - \varphi_{lim} \approx \pi/2$.

$$\sigma_{bottom\ not\ recollected} = \sigma_{bottom} \frac{2 \times \int_{\varphi=-\beta_{lim}}^{\beta_{lim}} \int_{\theta_{lim}}^{\pi/2} sin\theta cos\theta d\theta d\varphi}{\int_{\varphi=0}^{2\pi} \int_{0}^{\pi/2} sin\theta cos\theta d\theta d\varphi} \tag{6}$$

Finally, using the relations (1)-(2) and (6), the general expression (7) of the EEY depending on the geometrical parameters is established:

$$\sigma = \sigma_0 \frac{L}{L+d} + \sigma_0 \frac{d}{L+d} \frac{2}{\pi} \int_{\varphi=-\beta_{lim}}^{\beta_{lim}} 1 - sin^2 \left( \arctan \left( \frac{2h}{d} \frac{1}{\sqrt{1+tan^2\varphi}} \right) \right) d\varphi \qquad (7)$$

This expression clearly shows that the EEY only depends on the ratio of the dimensions *L, d* and *h*. As described in the previous section, a good agreement with the MC simulations from the code MicroElec is found when $\sigma_0 = 1.57$ is chosen. It corresponds to the experimental yield of a flat surface of silver. For L=d, the slope for $h/L \in [1; 10]$ and the plateau for $h > 10L$ are estimated correctly. The analytical expression shows that when the height of the structure increases, the number of electrons emitted from the bottom of the structures that are not recollected decreases to a corresponding yield of nearly 0.2. It results that the observed plateau is mostly due to the electrons emitted from the top of the structures ($\sigma_0/2$ when L=d). For a height h>10d, the analytical model shows the same slope as the MC simulations for values $L/d \in [0,5 ; 10]$. A plateau also appears for $d > 10L$. The analytical expression shows that when the width L is reduced, the yield $\sigma_{top}$ strongly decreases. The plateau observed on Fig. 6 is mostly due to the electron emitted from the bottom of the structures that are not recollected. Since the model overestimates these electrons, it appears that for small ratios $L/d$, the value of the plateau is overestimated.

## V.    STUDY OF CHECKERBOARD PATTERNS

Similarly to the study of grooves, in the case of checkerboard patterns, the structures are described by three parameters, illustrated in Fig. 8: the height h, the width *L* of the structure, and the width d between two neighboring structures. In the first part of this section, the effect of the height h of the structures is studied. In the second part the impact of the width L of the structures on the EEY is analyzed.

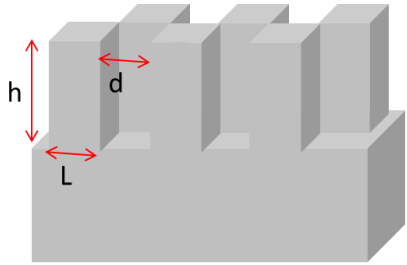

FIG. 8. Illustration of the checkerboard patterns used in this study.

## A.   HEIGHT OF THE STRUCTURES

To investigate the effect of the height h of the structures on the EEY, the two parameters L and d are kept constant and of the same value (*L=d*). Similarly to the study with the grooved patterns, an increase of the height of the structures leads to a reduction of the EEY. A similar result has been observed on the MC simulations of Ye *et al* [23] on copper and steel nanocones of dimensions between the nanometer and the micrometer. To evaluate more precisely this reduction, the maximum of the EEY is shown on Fig. 9. As can be seen, for a ratio h/L>10, the maximum of the EEY reaches a plateau with a value near 0.9. This value corresponds to a reduction of nearly 50 % of the EEY in comparison with the yield of a flat surface of silver. Similarly to the grooved patterns, the effect of checkerboard patterns is purely geometrical and can be compared to a simple analytical model (see section V.C) that corresponds to the black curve in Fig. 9.

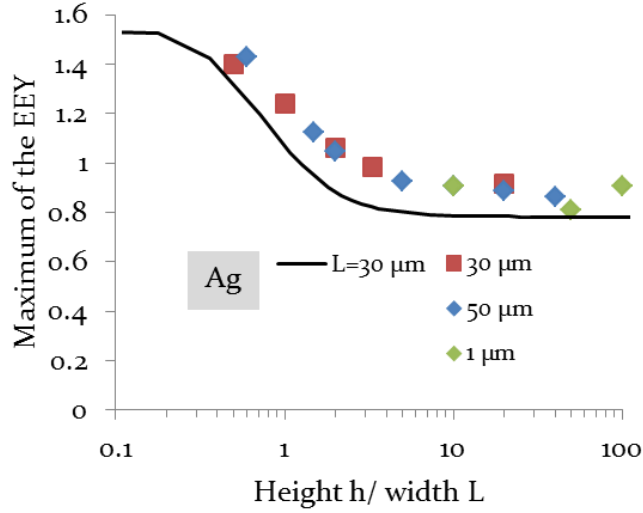

FIG. 9. Maximum of the EEY for a sample of silver with checkerboard patterns on its surface with a height h, a width L and a spacing d between two neighbouring structures (L=d). The maxima are calculated with an energy step of 50 or 100 eV for incident energies between 100 and 600 eV. The uncertainties are of less than 5 %. The black line corresponds to the analytical model described in section V.C.

In the case of checkerboard patterns, when the EEY decreases, as shown in Fig. 10a, the first crossover point $Ec_1$ is shifted to a higher energy. For example, for a ratio $h/L=2$ with $L=50\mu m$, a value of 384.23 eV is found. On the contrary, as shown in Fig. 10b, the second crossover point $Ec_2$ is shifted to a lower energy. For a ratio $h/L=2$ and $L=50\mu m$, a value of 497.70 eV is found. Taking into account the uncertainties of the MC simulations, a limit value of $h/L=5$ is determined for which no crossover is found. The analytical model (see section V.C) gives a limit value of 1.16.

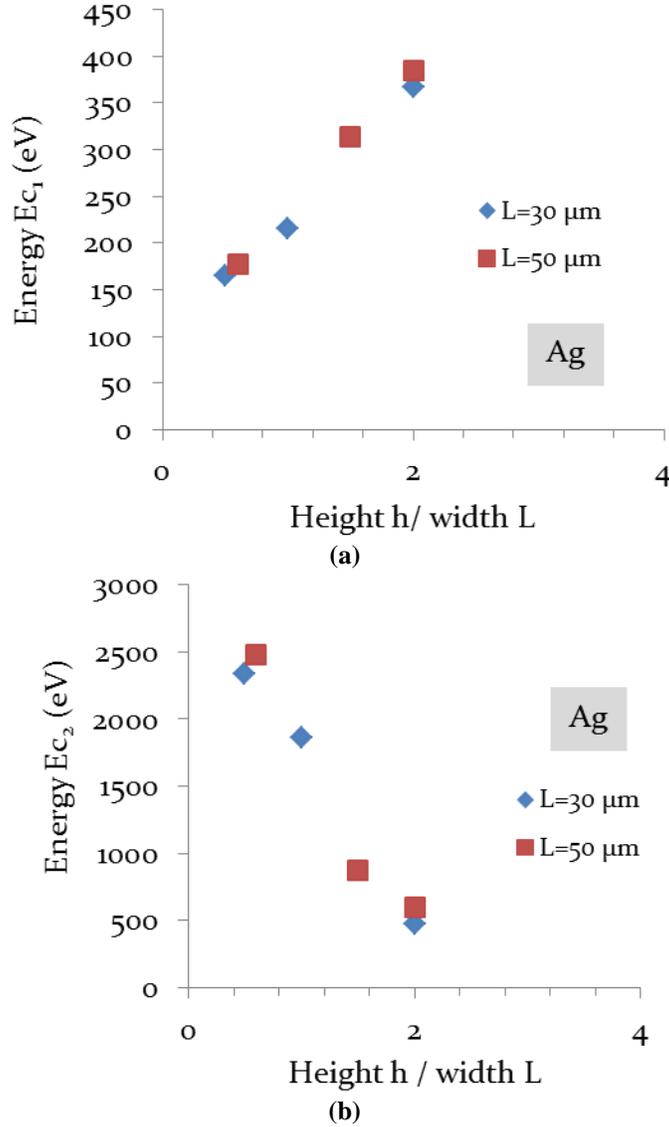

FIG. 10. Variation of the energy for the (a) first and (b) second crossover points in function of the ratio h/L with h the height of the and L the width of the structures in the case of chekerboard patterns. The energy of the crossover points are calculated by linear interpolation on the EEY. The EEY are obtained with a step of 50 or 100 eV between 100 and 600 eV and a step of 200 eV for higher energies.

## B. WIDTH OF THE STRUCTURES

To study the effect of the width L of the structures, the two parameters h and d are kept constant with $h \geq 10d$. For silver material, Fig. 11 shows the calculated maximum EEY for different values of height $h$ and width $d$. For a ratio $L/d=0.1$, the EEY reaches a value of 0.35. This corresponds to a reduction of the EEY of nearly 80 % compared to that of a flat sample. As can be seen on the figure, for a ratio $L/d$ smaller than 0.01, the EEY is reduced by nearly 98 % in silver material — the value for the maximum is $\sigma = 0.03$.. As the effect of roughness is purely geometrical, similar results are expected for structures with similar ratios but a different scale factor. This behavior is also depicted in Fig. 11 for d=30 μm ($h$=300 μm) and for $d$=1μm ($h$=10 μm).

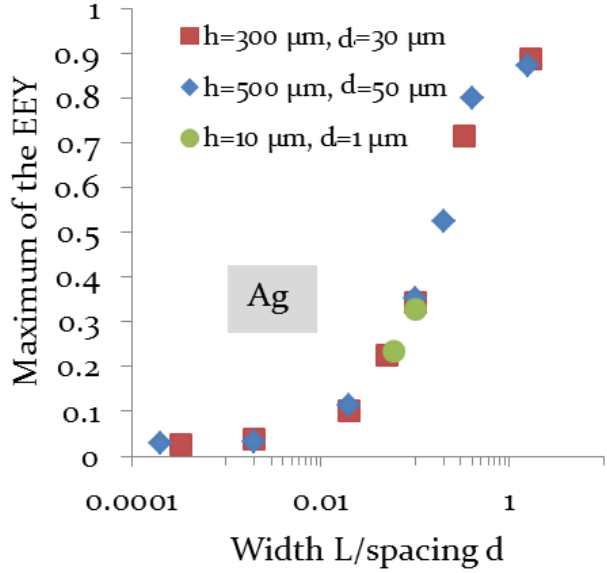

FIG. 11. Maximum of the EEY for a sample of silver with checkerboard patterns of different height h, widths L and spacing d. The values are calculated with an energy step of 50 or 100 eV for incident energies between 100 and 600 eV.

## C. ANALYTICAL MODEL

This section describes the analytical model presented in Fig. 10 in the case of checkerboard patterns with $L=d$. Similarly to the case of the grooves described in section IV. C, the effect of roughness is supposed to be identical regardless of the incident electron energy and the electrons that are emitted from the edges of the sides of the structures have been neglected. The yields $\sigma_{top}$ and $\sigma_{bottom}$ are then given by the relations (7) and (8).

$$\sigma_{top} = \sigma_0 \frac{L^2}{L^2 + d^2} = \frac{\sigma_0}{2} \qquad (8)$$

$$\sigma_{bottom} = \sigma_0 \frac{d^2}{L^2 + d^2} = \frac{\sigma_0}{2} \qquad (9)$$

In the case of checkerboard patterns, the number of electrons emitted from the bottom of the structures and then recollected shall be recollected. Fig. 12 shows a view from the top and from the front of the patterns. As can be seen on this figure, the electrons that are not recollected are those that reach the top of a hemisphere of radius $d/2=L/2$, i.e. the electrons that have an angle $\theta$ greater than a limit angle $\theta_{lim}$:

$$tan\theta_{lim} = \frac{h}{\sqrt{\left(\frac{d}{2}\right)^2 + \left(\frac{L}{2}\right)^2}} = \frac{\sqrt{2}h}{d} \qquad (10)$$

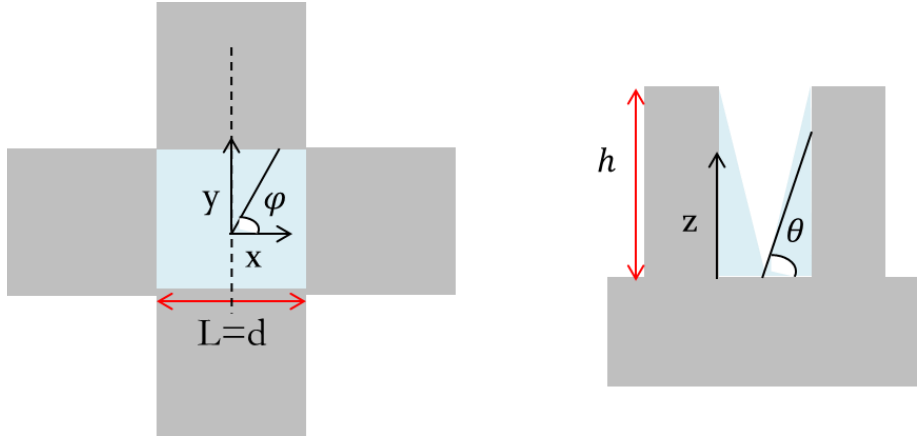

FIG. 12. Illustration of the checkerboard patterns used in the study. View from the top (left) and from the front (right). The electrons in the blue area are recollected by the structures.

Analogously to the grooved patterns, the electrons are supposed to be emitted according to a Lambertian angular distribution. In that case, the yield $\sigma_{bottom\ not\ recollected}$ is given by the equation (10):

$$\sigma_{bottom\ not\ recollected} = \sigma_{bottom}\ \frac{\int_{\varphi=0}^{2\pi}\int_{\theta lim}^{\pi/2} sin\theta cos\theta d\theta d\varphi}{\int_{\varphi=0}^{2\pi}\int_{0}^{\pi/2} sin\theta cos\theta d\theta d\varphi} \tag{11}$$

Hence, the general expression of the EEY for checkerboard patterns with $L=d$ is given by:

$$\sigma = \sigma_0\left[1 - \frac{1}{2}sin^2(\arctan(\frac{\sqrt{2}h}{d}))\right] \tag{12}$$

The principle of this calculation is comparable to the one of Sattler[10], but this model make a rough approximation on the definition of the limit angle. In our model the limit angle is supposed to be the same whatever the location of the emitted electron at the bottom of the pore, while in reality it depends on the distance of the emitted electron to the walls of the pore. For the "high" microstructures with a large aspect ratio $h/L$ ($L=d$), the simplification adopted here, limit the aperture angle and thus lead to an underestimation of the EEY as can be seen in Figure 9. However, as can be seen in Fig. 9, this analytical model is in fairly good agreement with the MC simulations. In Fig. 9, the EEY is calculated assuming $\sigma_0 = 1.57$. This value underestimates the EEY. A maximum of EEY around $\sigma_0 = 1.8$, provides a better agreement with Monte Carlo simulations. For $h/L$ in the range $[0,1;10]$ the slope of the curve is fairly well reproduced, and the plateau region is found for $h > 10L$. As can be noticed, the analytical expression shows that when the height of the structure is sufficiently high, the EEY tends to $\sigma = \sigma_0/2$. This value is explained by the fact that when the height increases (with L=d), most of the electrons emitted from the bottom of the structures are recollected by shadowing effects.

## VI.    DISCUSSION: COMPARISON OF GROOVED AND CHECKERBOARD PATTERNS

Fig. 13 and Fig. 14 show a comparison of the effect of grooves and checkboard patterns on the maximum of the EEY. Fig. 13 highlights the effect of the height h of the structures with *L=d*. Fig. 14 shows the effect of the width L of the structures when h≥10d.

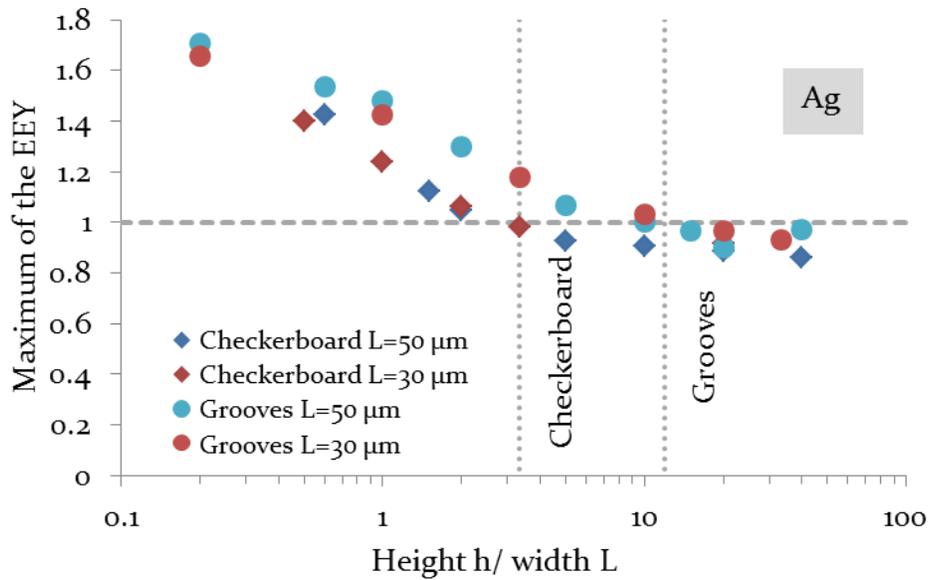

FIG. 13. Comparison of the maximum of the EEY for silver in the case of grooves and checkerboard patterns as a function of the ratio h/L with d=L=50µm and d=L=30µm.

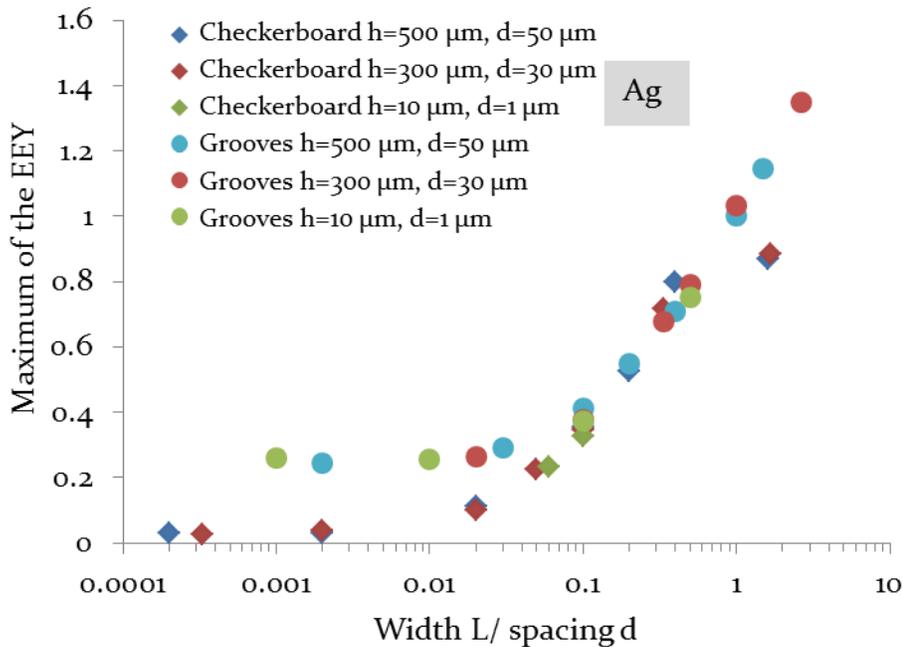

FIG. 14. Comparison of the maximum of the EEY for silver in the case of grooves and checkerboard patterns as a function of the ratio L/d with h>10d.

As can be seen, the checkerboard patterns give a reduction of the EEY greater than the grooved patterns. In Fig. 13, the EEY becomes lower than 1 for a ratio *h/L=3* for checkerboard patterns and for a ratio *h/L=15* for grooves. In Fig. 14, for a ratio $L/d \geq 0.1$ with h=10L, the maximum of the EEY has the same value for a surface of silver with grooves and with checkerboard patterns. However, for a ratio $L/d \leq 0.01$, the maximum of the EEY reaches a plateau which is much lower for the checkerboard patterns. This result was expected since there are fewer possibilities for the electrons to be recollected in the case of grooved patterns.

## VII.    COMPARISON WITH EXPERIMENTAL DATA

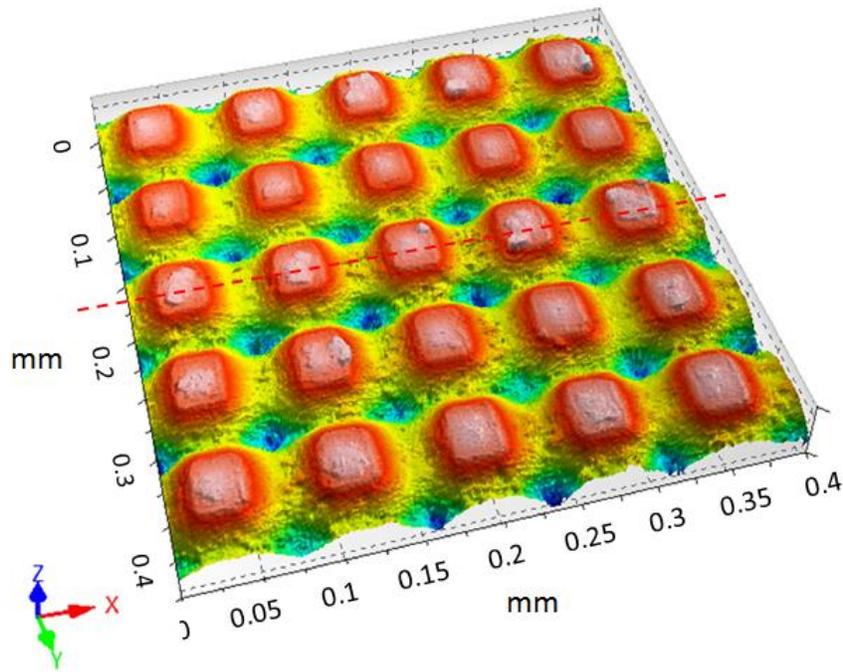

**(a)**

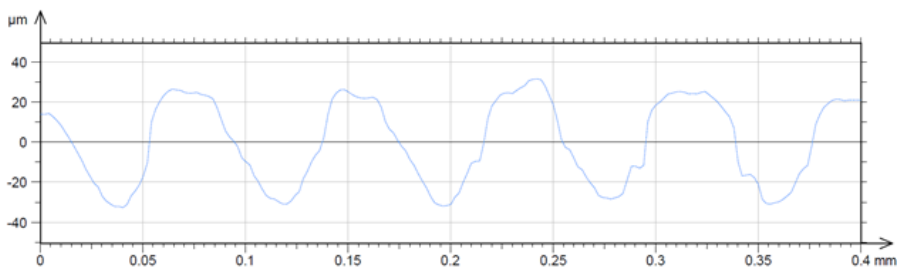

**(b)**

FIG. 15. Micro-structured Ag material irradiated with low energy electrons in the range [10eV, 1.5keV]. Checkerboard pattern with a height of ~60μm, 80μm periodicity. A cross section of the crenelated structure is given in (b). This corresponds to the dashed red line of the upper part of the figure (a).

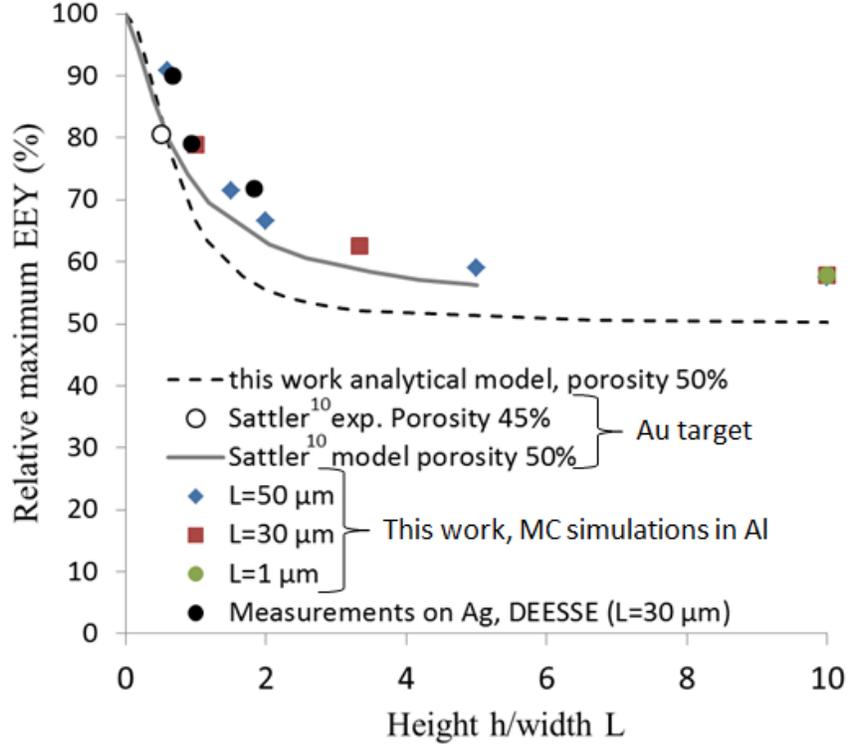

FIG. 16. Comparison with numerical simulations of measured relative maximum of the EEY for samples of silver etched with different checkerboard patterns with various aspect ratios h/L. In each case of simulation, for the different L values, d=L, that represents a 50% porosity ratio. Comparisons are made with the model of Satler[10] and an experimental measurement performed on a gold sample having a 50% porosity.

Some experimental irradiations have been performed in the DEESSE[4-6] facility introduced in the section III. Some aluminum samples having checkerboard structures have been irradiated with normal incident electrons with various energies in the range [10eV, 1.5 keV]. A morphological image of a sample is given in figure 15. We can see on this image that the crenellated pattern is made of structures having a height of ~60 μm with a periodicity of ~80 μm. The bottoms of the wells are less broad than the top of the structures. The comparison with the ideal modeled structures having perfect vertical walls is thus quite approximate, and for two reasons: firstly because the aspect ratio is not defined precisely for the real material, and secondly because the inclined walls of the real structures, are more efficient secondary electrons emitter than the vertical walls of the simulated structures. For that reason the comparison with the simulation results can only be relative. Three samples having different aspect ratio (h/L) have been tested and compared to simulation results. The periodicities of the three samples are similar and equal to nearly 80μm. The heights of the three samples are respectively 20 μm, 40μm and 60μm. The width L of the bottom of the structure have been estimated to be equal to around 30 μm giving aspect ratio respectively of 0.67, 0.93 and 1.83.

The EEY have been measured in the range [10 eV, 1.5 keV] and the maximum of EEY have been extracted and plotted as a function of the aspect ratio on the figure 16.

As the simulations are performed for "ideal" materials without taking into account any contamination nor oxidation, only relative maximum SEY have been compared. The SEY of reference correspond to the "flat" material that has not been etched with a checkerboard pattern. In each case of simulation, for the different $L$ values, the spacing $d$ between two patterns is taken equal to $L$ (the width of a pattern). If the porosity is defined as the ratio of the pore surface area to the total surface area, our simulations are representative of a material having a 50% porosity. That allows a comparison with both the model and experimental data of Sattler[10]. Sattler fabricated various micro-porous gold surfaces, having pores of various diameters, going from ~10µm up to ~40µm, the same order of magnitude of the dimensions used in our work. In the work of Sattler, the porosity was varied from 5% up to 75% while our porosity level is equal to 50%. In the case of the three samples tested in the DEESSE facility, the definition of both the porosity level and the aspect ratio is more debatable, limiting the scope of the comparison. The figure 16 presents this comparison. The comparisons are made with both the measurements performed within the DEESSE facility and the data of Sattler[10] (full line for model and open circle for experimental data). The measurement has been obtained by Sattler[10] on a gold sample having a porosity of 45% and an aspect ratio of 0.5. The analytical model described in this work is also shown in the figure 15 (dashed line).

First, it should be noted the quite good agreement between our simulations and our experimental data in view of the differences between the real and the simulated geometrical structures. The relative decrease of the EEY as a function of aspect ratio is in the same order of magnitude for both experiments and Monte Carlo simulations. But, the differences between the simulated structures and those used experimentally do not permit to draw more precise conclusions. The analytical models provide relatively good results too. Both models predict a decreasing SEY as a function of the aspect ratio within the right order of magnitude. For big aspect ratios (>2) the analytical model of Sattler is very close to our Monte Carlo simulations (~5% difference for $h/L=5$). But, our analytical model underestimates the EEY due to the used approximation that underestimates the aperture angle of the pore (cf. section V-C). For smaller aspect ratios the two models become closer and closer. They differs slightly (~13% difference for an aspect ratio of 0.6) with the Monte Carlo simulations. The experimental measurement of Sattler[10], obtained on a gold sample with an aspect ratio of 0.5 and a porosity of 45 % is quite close to the Sattler model but differs from our Monte Carlo simulations (~13%).

The underestimation of the SEY by the analytical models could be due to the assumption that supposes that the electrons that impact the pore sidewall are captured. That means that only the first generation of secondary electrons is considered in the analytical models. Taking into account the second generation of secondary electrons in the Monte Carlo simulations contributes to enhance the SEY in comparison to the analytical models. But, that cannot explain the relatively low EEY measured by Sattler on his porous surface. We must keep in mind that any surfaces treated with a chemical etching, a laser ablation, etc…will present some nanoscale topographical variations that will have a direct impact on the measured EEY. Depending on the treatment it can enhance or reduce the SEY, and it is thus difficult to expect a better accuracy as the ~13% observed here.

## VIII.    CONCLUSION

The low energy electromagnetic model of GEANT4 (MicroElec) has been extended down to a few eV. After comparing, in a previous work [18, 19], the EEY obtained from this new version with experimental measurements [4, 5], 3D simulations have been performed to study the impact of surface morphology on the low energy electron emission. Grooved and checkerboard patterns of different dimensions have been studied. These simple patterns depend only on three shape parameters: the height h, the width L of the structure, and the spacing d between two neighboring structures. The EEY decreases for increasing height values. It also decreases according to the width of the structures present on the surface (L). These simulations have shown that when the height of the roughness structures reaches 10 times the lateral dimensions (L and d) the EEY tends to a lower limit. In that case the EEY is divided by two compared to that of a flat sample. While having a sufficiently high height for the structure, the decrease in width of the structures has two different effects: it flattens the overall shape of the EEY and considerably reduces the maximum value of the EEY. A combination of an increase of the height with a decrease of the distance between two roughness structures can lead to a reduction of the EEY of almost 80 % for grooved patterns, and of almost 98 % for checkerboard patterns, compared to that of a flat sample. This purely geometric effect leads to similar results for aluminum and silver target materials. A simple analytical model, capable to reproduce the effect on the EEY of checkerboard and grooved patterns, have been proposed. This model is found to be in good agreement with the Monte Carlo simulations.